\documentclass[twocolumn,showpacs,aps,prd,amsfonts,
superscriptaddress,nofootinbib]{revtex4-1}
\usepackage{graphicx}
\begin{document}
%


\newcommand*{\zilina}{Physics Department, University of \v{Z}ilina, Slovakia}
\newcommand*{\praha}{Institute of Experimental and Applied Physics,
Czech Technical University in Prague, Czech Republic}

\title{A brief outline of the top-BESS model}

\author{Mikul\'{a}\v{s} Gintner}
\affiliation{\zilina}
\affiliation{\praha}
\author{Josef Jur\'{a}\v{n}}
\affiliation{\praha}
\author{Ivan Melo}
\affiliation{\zilina}

\maketitle

Despite the great success of the Standard Model (SM) one essential
component of the theory remains a puzzle: it is the actual
mechanism behind the electroweak symmetry breaking (ESB).
Spontaneous breaking of electroweak (EW) symmetry accompanied by the
Higgs mechanism is the way to reconcile the massive gauge bosons
with the principle of gauge invariance.
A direct consequence of this hypothesis is the
presence of the scalar Higgs boson in the particle spectrum, not
observed as of yet, though.

There is a host of candidates for alternative extensions of the SM
that offer their own mechanisms of ESB. The formalism of effective
Lagrangians can be used to classify and investigate their
phenomenology at energies accessible at the LHC and other near
future colliders (ILC). We have introduced the {\it top-BESS
model} (tBESS)~[1] as an effective description of a high-energy
extension of the Higgsless SM where new strong interactions are
responsible for ESB. The full formulation of the tBESS model can
be found in~[1]. In the following we would like to present its
basic properties.


The tBESS is the modified version of the BESS (Breaking EW
Symmetry Strongly) model~[2]. Both models describe a new $SU(2)$
vector boson triplet that can represent the spin-1 bound states of
hypothetical new strong interactions. They are based on the
$SU(2)_L\times SU(2)_R\times U(1)_{B-L}\times SU(2)_{HLS}$ global
symmetry of which the $SU(2)_L\times U(1)_Y\times SU(2)_{HLS}$
subgroup is also a local symmetry. "HLS" stands for the
\textit{hidden local symmetry}~[3] which is an auxiliary gauge
symmetry introduced to accommodate the $SU(2)$ triplet of vector
resonances. Beside the triplet, the models contain only the
observed SM particles.

In the gauge sector new physics is parameterized by the $SU(2)_{HLS}$
gauge coupling $g''$ and another parameter $\alpha$.
In the limit when $g$ and $g'$ are
negligible compared to $g''$ the masses of the neutral and
charged resonances are degenerate, $M_{V} = \sqrt{\alpha}g''v/2$.
Higher order corrections in $g/g''$ result in the
mass splitting such that $M_{V^0}>M_{V^\pm}$.

In the tBESS model we modify the \textit{direct} interactions of
the vector triplet with fermions. While in the BESS model there is
a universal direct coupling of the triplet to all fermions of a
given chirality, in our modification we admit direct couplings of
the new triplet-to-top and bottom quarks only. Our modification is
inspired by the speculations about a special role of the top quark
in the mechanism of ESB.

In the tBESS model, the triplet-to-top/bottom coupling is
proportional to the $SU(2)_{HLS}$ gauge coupling $g''$
multiplied by the $b_L$ and $b_R$ parameters for the left
and right fermion doublets, respectively. In addition, we have
disentangled the triplet-to-top-quark right coupling from the
triplet-to-bottom-quark right coupling by introducing
a free parameter $p$, $0\leq p\leq 1$. When $p<1$,
the strength of the triplet-to-bottom right coupling is weakened.
The $SU(2)_L$ symmetry does not allow the same
splitting for the left quark doublet.

The BESS/tBESS symmetry admits two more invariant terms which were
not considered by the authors of the BESS model~[2]. The terms
introduce additional free parameters, $\lambda_L$ and $\lambda_R$.
They do not have a significant impact on the behavior of the model
at energies around the mass of the vector triplet, but they do
modify the couplings of the EW gauge bosons with fermions.

New vector resonances mix with the EW gauge bosons. It results in
the \textit{indirect} --- mixing-induced --- interactions of
the vector triplet with fermions. For the light fermions this is the
only way they can interact with the vector triplet in the tBESS
model. Of course, the interactions are suppressed.

\begin{figure}[h]
\begin{center}
\includegraphics[width=40mm]{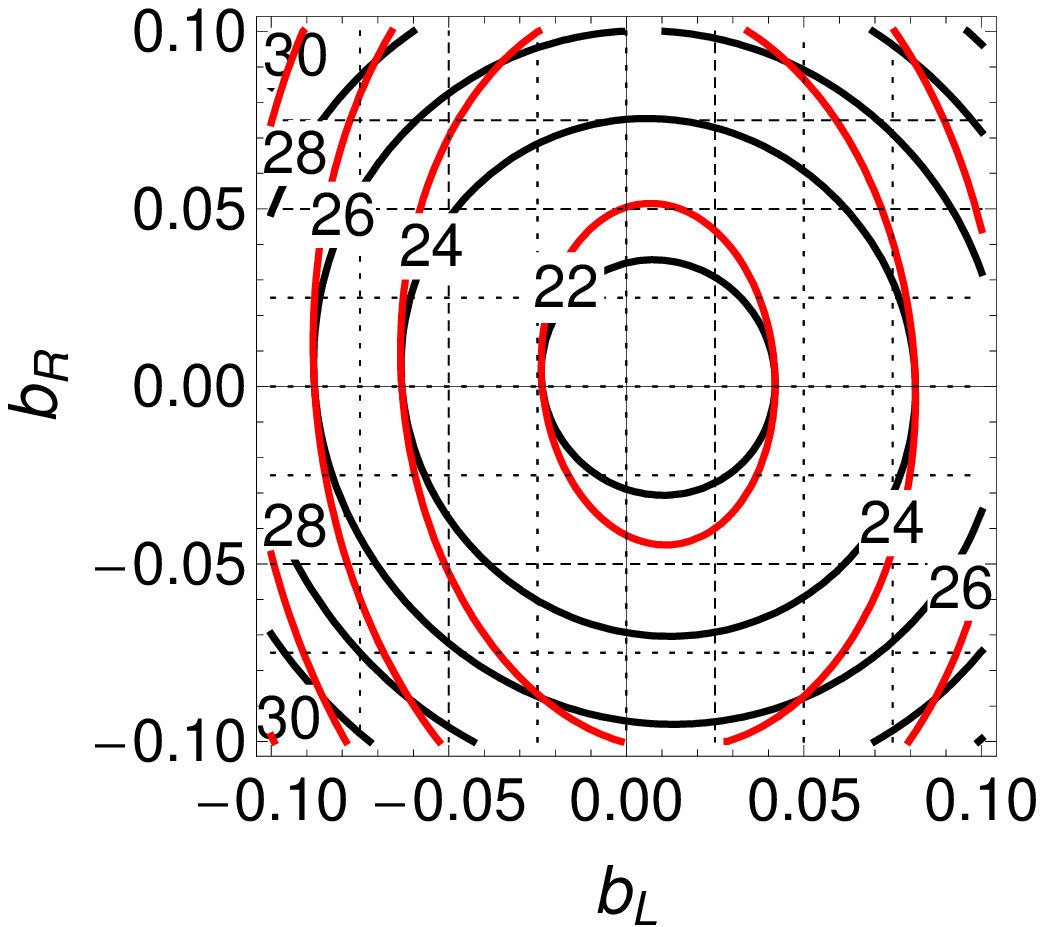}
\includegraphics[width=40mm]{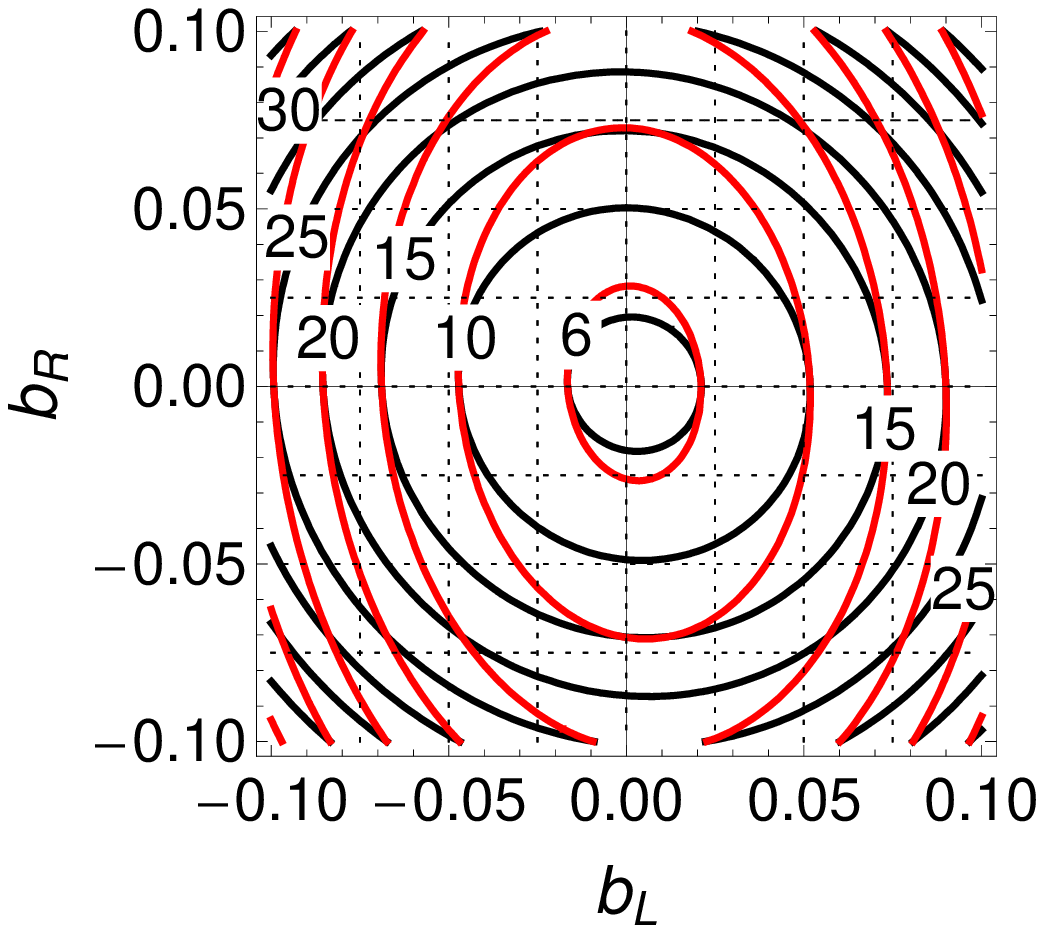}\\[0.3cm]
\includegraphics[width=40mm]{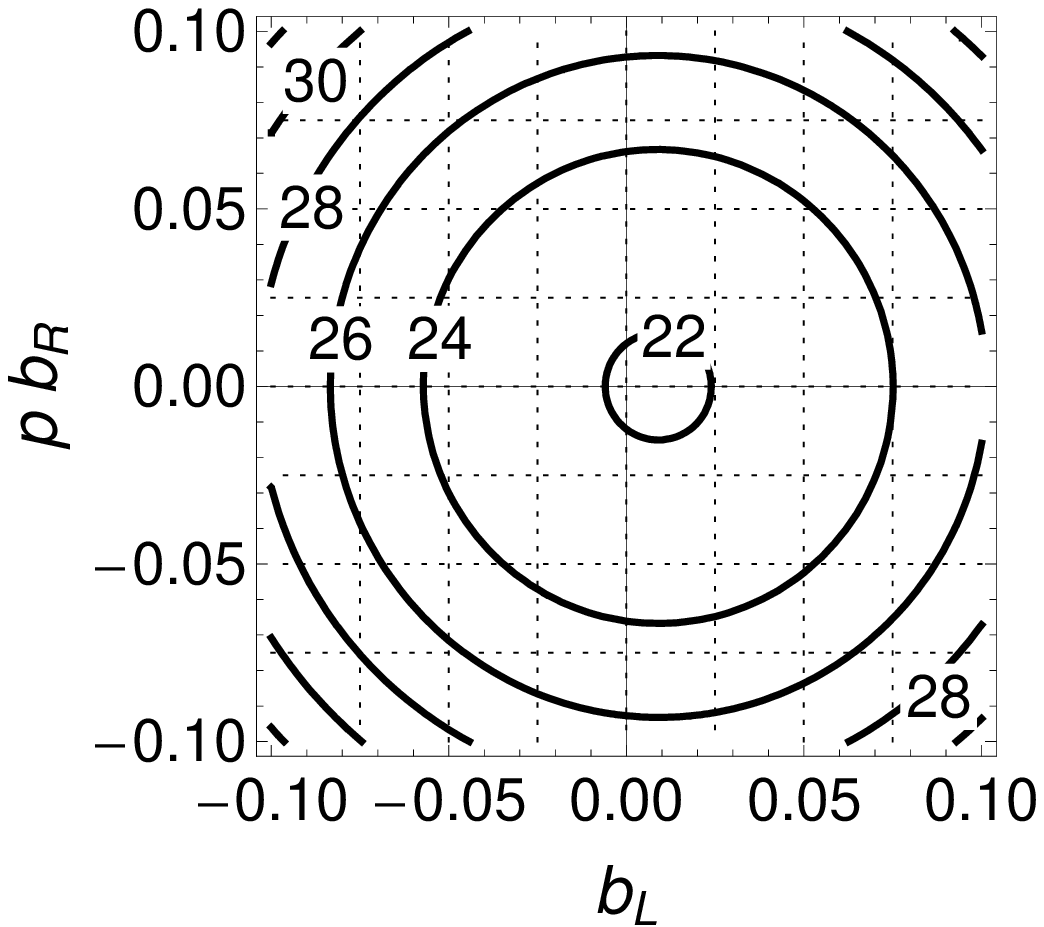}
\includegraphics[width=40mm]{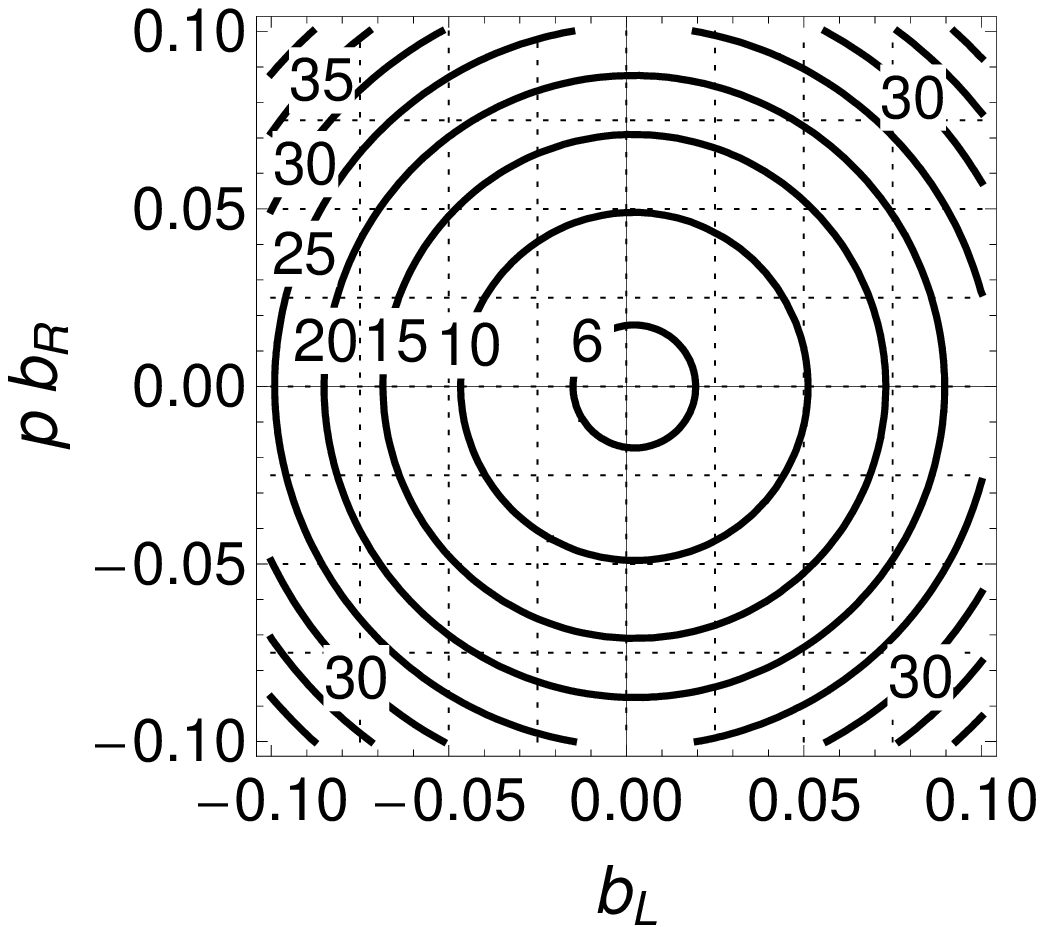}
\end{center}
\vspace{-2mm} \caption{\label{fig:DWcontours}
 The total decay width (labels in GeV) contours of the tBESS triplet.
 The $V^0$ decay (upper row) for the cases of $p=1$ (black)
 and $p=0$ (red) and the $V^\pm$ decay (bottom row).
 The left and right columns correspond to
 $g''=10$ and $g''=20$, respectively. All graphs have been
 plotted for $M_{V^0}=1$~TeV and $\lambda_L=\lambda_R=0$.}
\end{figure}

The vector triplet predominantly decays to the electroweak gauge
bosons, $W^\pm$ and $Z$, and/or to the third generation of quarks
(except a special case discussed below). The quark decay channels
prevail when the moduli of $b$ parameters assume sufficiently
large values. A typical decay width of the 1~TeV triplet is about
a few tens of GeV. The total decay widths of the tBESS resonances
are shown in Fig.~\ref{fig:DWcontours}.
%

The interplay of the direct and indirect couplings can diminish
or even zero a particular top/bottom quark channel decay width
of the vector resonance for some nonzero values of the $b$ parameters.
Thus, it might happen that even though the direct couplings
are nontrivial the resonance will not decay through the given top
and/or bottom channel. We call the area of the parameter space
where the decay width of the resonance is lower than its indirect coupling
generated value, the \textit{Death Valley} (DV).
The DV does not virtually depend on the $\lambda$ values.

\begin{figure}[h]
\begin{center}
\includegraphics[width=80mm]{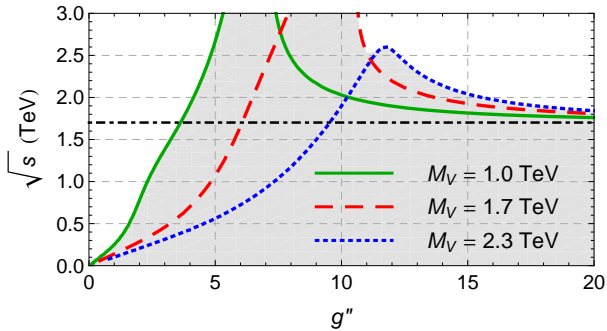}
\end{center}
\vspace{-2mm} \caption{\label{fig:Ulimit}
 The tree-level unitarity constraints for various masses of the vector triplet:
 $M_V=1$~TeV (solid line), $1.7$~TeV (dashed), $2.3$~TeV (dotted).
 The horizontal dashed-dotted line is the Higgsless SM unitarity limit of 1.7~TeV.
 The shaded area indicates the region where the unitarity holds.
 No couplings to fermions are assumed.}
\end{figure}

The tBESS model is nonrenormalizable and violates unitarity
at some energy. The tree-level unitarity constraints have been
obtained using the Equivalence theorem (ET) approximation
of the $W_L^+W_L^-$, $Z_L Z_L$, $W_L^\pm Z_L$, and $W_L^\pm W_L^\pm$
scattering, see Fig.~\ref{fig:Ulimit}. The nonrenormalizability
implies the upper limit, $\sqrt{s}\approx 3$~TeV, on the applicability of the ET.
%

\begin{figure}[h,t]
\begin{center}
\includegraphics[width=80mm]{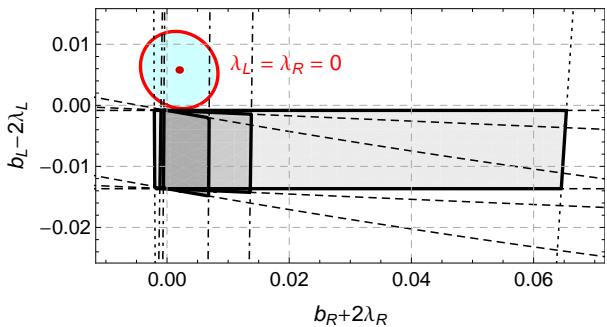}
\end{center}
\vspace{-2mm} \caption{\label{fig:LElimit}
 The intersections (shaded areas) of the $90\%$~C.L. allowed regions
 for $p=0$ (the lightest gray), $p=0.5$ (middle gray), and $p=1$ (the darkest gray)
 when $M_{V^0}=1$~TeV and $g''=13$. The red circle bounds
 the $V^0\rightarrow t\bar{t}$ DV region when $\lambda_{L,R}=0$.
 The red dot indicates the zero value of the decay width.}
\end{figure}

The existing \textit{electroweak precision data} (EWPD) restrict
tBESS induced deviations from the SM at relevant energies. We have
used the measured values of the $\epsilon_{1,3,b}$~[4] parameters
along with the measurement of the $B\rightarrow X_s\gamma$ and
$Z\rightarrow b\bar{b}$ decays, and the D0 measurement of
$p\bar{p}\rightarrow WZX$. It resulted in the \textit{low-energy
limits} on the parameters of the tBESS model. In particular, the
EWPD restrict $b_L-2\lambda_L$ and $b_R+2\lambda_R$ rather than
$b$'s and $\lambda$'s individually. The intersection of the
allowed regions for $M_{V^0}=1$~TeV and $g''=13$ is shown in
Fig.~\ref{fig:LElimit}.
%
In principle, $b$'s and $\lambda$'s can assume \textit{any} values
if their difference/sum falls within the allowed interval. However, the
greater their values, the finer tuning is necessary. Even with
this qualification, the low-energy limits on tBESS parameters are
significantly less restrictive than in the BESS case.

In Fig.~\ref{fig:LElimit}, the DV for $V^0\rightarrow t\bar{t}$ decay
is also shown. In the $(b_L-2\lambda_L,b_R+2\lambda_R)$ space,
the position of the DV depends on $\lambda$'s.
The displayed case corresponds to $\lambda_L=\lambda_R=0$.
The size of the DV region shrinks when $g''$ grows.

\begin{figure}[h,t]
\begin{center}
\includegraphics[width=40mm]{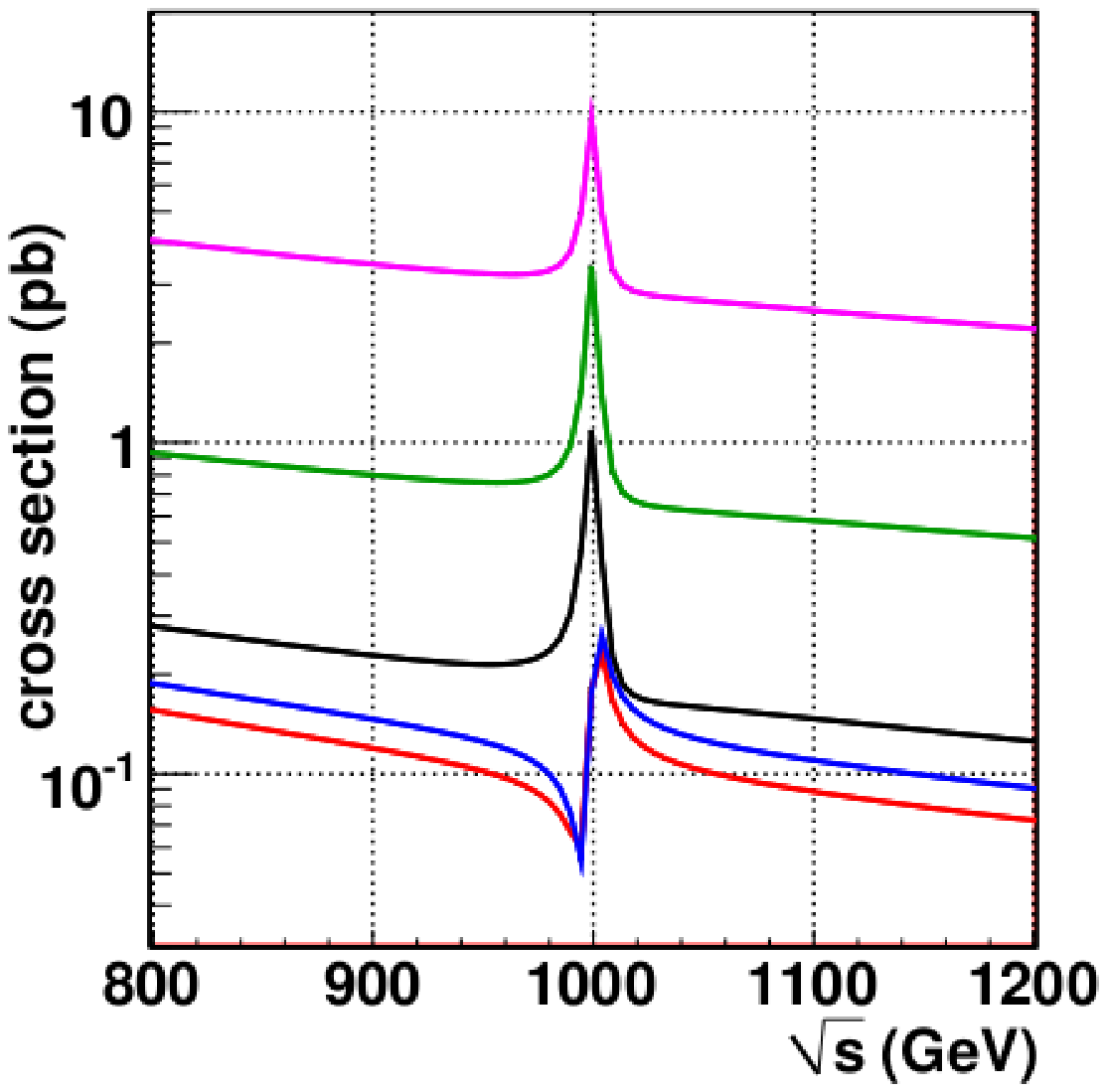}
\includegraphics[width=40mm]{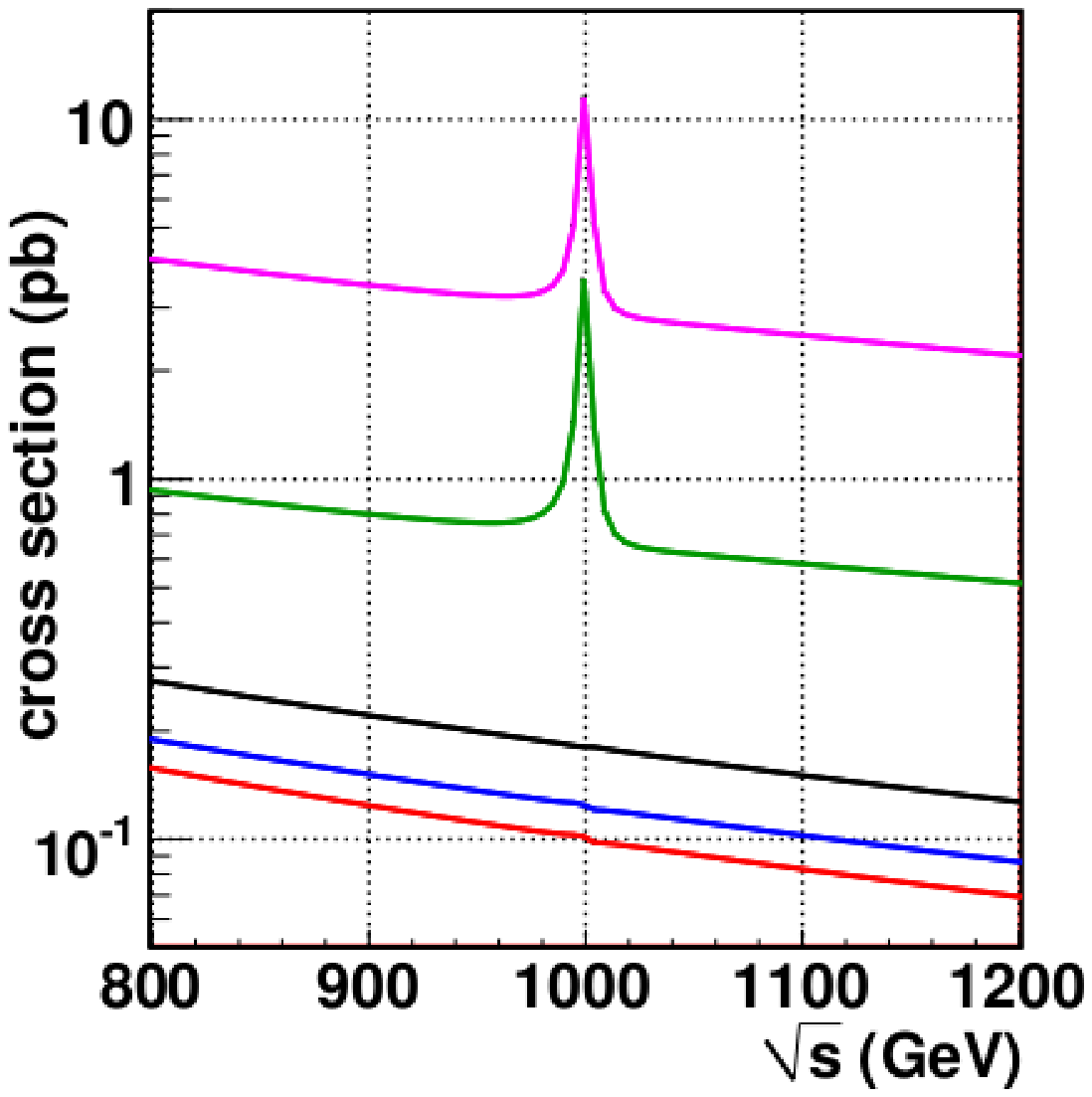}
\end{center}
\vspace{-2mm} \caption{\label{fig:Xsection}
 The cross sections of $e^-e^+\rightarrow W^+W^-$ (magenta),
 $u\bar{d}\rightarrow W^+Z$ (green), $e^-e^+\rightarrow t\bar{t}$ (black),
 $u\bar{d}\rightarrow t\bar{b}$ (blue), $e^-e^+\rightarrow b\bar{b}$ (red);
 $M_{V^0}=1$~TeV and $g''=20$.
 Values of the fermion parameters are chosen far away from the DV's (left)
 and at the bottom of the DV's (right) of all three top/bottom channels.}
\end{figure}

The DV effect can hide signals expected in scattering
processes. Even if the tBESS resonances existed and coupled to the
third quark generation there would be no peak in the
scattering experiments for certain final states with top
and/or bottom quarks. This would occur if the parameters happened
to have their values inside the DV region. To illustrate the DV
effect on the scattering amplitudes we have plotted the cross
sections for five processes in Fig.~\ref{fig:Xsection}.
%

As a final note, our calculations suggest that there are allowed values of the
tBESS parameters which can result in detectable signals at the LHC
and/or the ILC. However, this is far from conclusive and a deeper systematic
study would be required to settle this question.
\\


\noindent ACKNOWLEDGMENT: The work of M.G. and J.J. was supported
by the Research Program MSM6840770029 and by the project
International Cooperation ATLAS-CERN of the Ministry of Education,
Youth and Sports of the Czech Republic. J.J. was also supported by
the National Scholarship Program of the Slovak Republic. M.G. and
I.M. were supported by the Slovak CERN Fund.\\

\noindent REFERENCES \\
$[1]$ M.~Gintner, J.~Jur\'{a}\v{n}, I.~Melo, Phys. Rev. D
\textbf{84}, 035013 (2011). \\
$[2]$ R.~Casalbuoni, S.~De~Curtis, D.~Dominici, and R.~Gatto,
Phys. Lett. \textbf{155B}, 95 (1985); Nucl. Phys. \textbf{B282},
235 (1987). \\
$[3]$ M.~Bando, T.~Kugo, and K.~Yamawaki, Phys. Rep. \textbf{164},
217 (1988). \\
$[4]$ G.~Altarelli, R.~Barbieri, and F.~Caravaglios,  Nucl. Phys.
\textbf{B405}, 3 (1993).

%
%

\end{document}